\documentclass[a4paper, conference, 10pt]{IEEEtran}
\usepackage{verbatim}
\usepackage[dvips]{graphics} 
\usepackage{graphicx}
\usepackage{times}
\usepackage{epsfig}
\usepackage{latexsym}
\usepackage{amsmath}
\usepackage{cite}
\usepackage{textcomp}
\usepackage{enumerate}

\usepackage{authblk}\usepackage{algorithmic,algorithm}
\begin{document}

%\title{Novel Delay Tolerant Relay Communication For Rural Agricultural Participatory Sensing}
\title{A Revolutionary Rural Agricultural Participatory Sensing Approach Using Delay Tolerant Networks} 
\author{Prateek Gupta, Bhushan Jagyasi, Bighnaraj Panigrahi, Hemant Kumar Rath, Srinivasu Pappula, Anantha Simha}
\affil{TCS Research India, \\Email:\{prateek.gupta5, bhushan.jagyasi, bighnaraj.panigrahi, hemant.rath, srinivasu.p, anantha.simha\}@tcs.com}
\maketitle

\begin{abstract}
To provide sustainable digital agro-advisory services to farmers, seamless flow of information from the farmers to the experts/expert systems, and vice versa is required. The query generated by the farmers, which may contain multimedia data regarding disease or pest attack in the crops is required to be transmitted to the experts for analysis. Further, after analyzing the query, an alert or advice from the expert system is required to be communicated back to the farmers within some tolerable delay. However, in a country like India, network connectivity is extremely poor in several agricultural regions which makes the end-to-end connectivity between the farmers and the expert system intermittent. Therefore, providing agro-advisory services to farmers in a reasonable time becomes a challenge. In this paper, we propose a Delay Tolerant Network (DTN) based relay application model which enables agro-advisory services to farmers located in \emph{No-network} or \emph{Poor-network} zones. In the proposed model, end-to-end communication has been enabled with the help of Device to Device (D2D) communication and by introducing mobile relay nodes which can carry the queries (responses), from (to) the poor or no network zones to (from) the zones where communication is possible. Implementation of this model has been presented for the tea farmers of West Bengal and Assam, which can be extended for various other applications in future.

\end{abstract}

\section{Introduction}\label{sec:intro}
In India, cellular network connectivity is very poor in several villages. It has been observed that rural and tribal folks travel for several kilometers to make a voice call. Data connectivity of 2G/3G is even worse in these areas. From our extensive involvement in \emph{Digital Farming} initiatives \cite{srini_csi}, the communication barrier due to the poor cellular network connectivity has been felt in several deployment locations. For instance, in remotely located tea farms and tea estates of Assam, telecommunication network is intermittent and very poor. Further, in the Araku constituency of Andhra Pradesh, which is known for its organic cultivation of coffee, rubber and spices, network is only available in selected pockets. Due to this, farmers of these regions face difficulty to get the relevant advice on their cultivation queries.
 
\begin{figure}[h!]
\centering 
\includegraphics[width=0.49\textwidth]{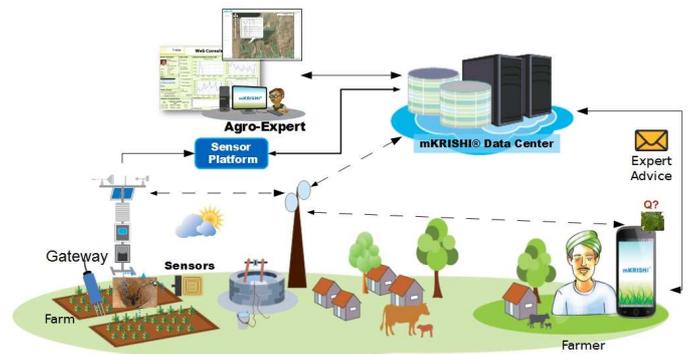}
\caption{Digital Farming Platform}  
\label{digital_framework}\vspace{-1mm}
\end{figure}

In the recent past, we have seen increasing applications of the Human Participatory Sensing in urban scenarios \cite{burke06, campbell06, Blaschke2011, bhushan_csi} such as traffic control, disaster information flow, epidemic monitoring, etc. In that direction, we have proposed a Rural Participatory Sensing (\emph{RuralSense}) framework for agricultural applications \cite{Jayant_Percom2015} and have deployed the same for tea crop in West Bengal and Assam. A RuralSense mobile application is provided to small and marginal tea growers of these areas in order to facilitate them to report various geo-tagged events from their farms. In addition, access to the web-based dashboard has also been provided to the experts to analyze the reported events and to advise back to the growers. These events serve multiple purposes such as (a) asking queries to the experts, (b) digitization of the farm diaries, (c) self certification or tracing of the chemical applications and (d) generating contextual data for developing pest/disease/ yield/ water models for the cultivated crops. Fig. \ref{digital_framework} illustrates a model digital farming platform; dotted lines indicate actual/physical communication between the farmers and the expert system either through cellular or any other communication mode. The continuous lines indicate the logical flow of messages between the sensors placed in the farm and the expert system. This framework for connecting farmers with the experts warrants a good communication network, which is the main focus of this paper. To overcome the limitations of poor cellular networks as discussed before, we propose {\emph{DTN-RuralSense}}, a novel Delay Tolerant Network (DTN) approach for agriculture applications. Note that, throughout this paper we have used query and event interchangeably.

Delay Tolerant Networks \cite{fall2003} has gained the interest of many researchers in the last decade. It is designed for harsh and challenging conditions where continuous Internet connectivity can not be guaranteed. To compensate data loss due to frequent disconnections, a relay node (DTN node) is required to store data packets for long time periods until connection to a forwarding node is established again. The use of this paradigm becomes critical in challenging scenarios like satellite, military, and rural applications. In countries like India, where the lack of infrastructure makes seamless end-to-end connectivity difficult, DTN can be used as an alternative.
  
In this paper, we have introduced a novel DTN-based agro-advisory framework to farmers working in a remote location with limited network connectivity. Using DTN framework, we have proposed a mobile relay-based architecture which can be deployed in remote villages and provide expert advice to farmers' queries in a time bound manner. We have also provided an Android-based deployment framework to realize this framework. The rest of the paper is organized as follows. In Section \ref{s:relWork}, we analyze the related work and in Section \ref{s:system_model}, we explain the problem and our approach towards a practical solution. We then discuss the deployment framework in Section \ref{s:deployment}, and conclude this paper in Section \ref{s:conclusion}.

\section{Related Works}\label{s:relWork}
Participatory Sensing \cite{burke06, campbell06, Blaschke2011} has gained tremendous growth in the recent past and has created various innovative applications using crowd-sensing and crowd-sourcing technologies. In these technologies, huge data from the crowd workers is generated and processed for appropriate actions. We have explored crowd-sensing in our previous work on Rural Participatory Sensing \cite{Jayant_Percom2015} and Distributed Crowd Sourcing \cite{Singh_Indicon2014}. In \cite{Jayant_Percom2015}, events are captured and reported with the help of a mobile application while in  \cite{Singh_Indicon2014}, tasks are assigned to the farmers to diagnose the plant diseases and pest classification. It is believed that for rural applications, crowd assisted sensing is an important source of data generation and the analytics performed on this big data shall serve the society for the years to come. 

DTN was initially intended for Inter Planetary Networks (IPNs) \cite{akyildiz2003} with a low network dynamic of satellites and rovers. In the DTN paradigm, communications can be intermittent and hence the nodes need to store the data till connection to a forwarding node is available again. By doing this, data communication can be guaranteed, even though the delay is substantial. In mobile wireless networks, the applications of DTNs have been envisioned for Vehicular Ad-hoc NETworks (VANETs) \cite{pereira2012}, Under Water Networks (UWNs) \cite{small2003}, Pocket Switched Networks (PSNs) \cite{hui2005}, and suburb networks for developing regions \cite{pentland2004}, etc. 

To provide Internet connectivity in rural India a Sustainable Access in Rural India (SARI) program \cite{akyildiz2003}, a DTN like framework has been initiated in which Internet kiosks have been distributed in different villages. However, providing and maintaining kiosks in every village is not feasible. Therefore, another project called Computers on Wheels (COW) \cite{conroy2006} was started in 2006, in which motorcycles with Internet equipment act as mobile kiosks and travel to remote villages to collect the data from users. In this paper, we explore the feasibility of the DTNs in enabling value added services to the rural farmers in a time bound manner. We use the smartphones/mobiles with the mKRISHI{\textsuperscript{\textregistered} application instead of kiosks and provide DTN-based end-to-end connectivity to the farmers.

\section{Problem Statement and Solution Approach} \label{s:system_model}
Time-bound agro-advisory services to farmers located in a remote area with either no or limited network connectivity is the challenge considered in this paper. The key goal of this paper is to remove the communication barriers due to unavailability of cellular networks.

In one of our endeavors on Rural Participatory Sensing (RuralSense) framework, farmers can collect (sense) various farming related information such as disease, pest, nutrient disorder and other activities. mKRISHI{\textsuperscript{\textregistered} is a Personalized Services Delivery Platform that enables two-way information exchange between farmers and expert systems that include virtual knowledge banks, Agriculture Experts and Procurement Officers (PO), etc. At the farmers' end, it uses participatory sensing (RuralSense) to collect and digitize the field data and mobile-based framework for the data communication. Each event/query generated at the farmers' end may consist of geo-tagged photographs, voice clip, meta-data, etc., and has a unique pair of Event ID (EID) bounded with the User ID (UID) which is assigned from the server. Moreover, it uses an application which can store the captured events/query in the local memory for communication at a later point of time.

At the expert system's end, it uses analytic tools and expert advisory services to evaluate the sensed data obtained from the farmers before communicating actionable advice to the farmers. This enables digitization of the farm and the farming related events. Note that, the availability of this information is extremely critical for decision makers in responding to the farmers' queries, in routing their agriculture inputs to the right location and at the right time. In poor network zones, farmers are however finding it difficult to have a seamless communication of the information. Therefore, a solution is desired which can be deployed in the mKRISHI{\textsuperscript{\textregistered} \cite{srini_csi} framework such that seamless time-bounded communication can be realized. Moreover, the new solution should be easily integratable with the existing mKRISHI{\textsuperscript{\textregistered} framework. It should also be easy to use for the farmers.
 
\subsection{Our Solution - a DTN Approach}
In this section, we propose a novel solution which can be included in the mKRISHI{\textsuperscript{\textregistered} framework without much changes at the end points. The aim here is to provide network connectivity to the farmers in an alternative way. To realize this, we propose to use relay-based communication which can operate on the principle of DTN; an opportunistic mode that becomes critical in challenging scenarios like satellite applications and rural communications in emerging countries like India or Africa, where the lack of an infrastructure makes regular data communications almost impossible. DTN communications are thus the natural choice for a networking paradigm where nodes can be disconnected from the regular network for the majority of the time and exchange of data can take long time. To provide DTN-based solution, we propose a new entity into the mKRISHI{\textsuperscript{\textregistered} framework called \emph{Relay Node}. A relay node can be any other smartphone which can be used by other 
fellow farmers, farm agents, or field executives, that comes in close proximity to the farmers at the remote locations and has the following capabilities: (i) collect the farmers' sensed data or queries using Device to Device (D2D) services, (ii) transmit the collected queries from one or more farmers to the expert system using available network technologies (Wireless Fidelity - WiFi, cellular or any other) in a time-bound manner, (iii) collect and communicate the acknowledgement (ACK) and advisories from the expert system to the farmers. Fig. \ref{soln_DTN} illustrates the possible modes where multi-hop relay communication can be introduced to eliminate the network connectivity problem at the farmers' end. 

\begin{figure}[h!]
\centering 
\includegraphics[width=0.49\textwidth]{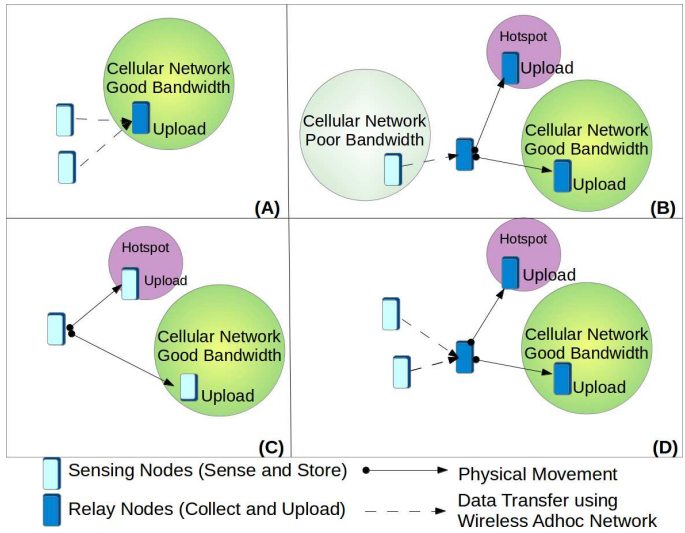}
\caption{Problem Cases - Need for DTN: (A) No network farmer zone, Relay with network, (B) Poor network farmer zone, Relay movement to a network zone, (C) No network farmer zone, farmer movement to a network zone, (D) No network farmer zone, Relay movement to a network zone}  
\label{soln_DTN}\vspace{-3mm}
\end{figure}

\subsection{Building Blocks of our Solution}
Various modules related to our architecture and the communication possibilities are explained through Fig. \ref{soln_BB}. Basic building blocks of our solution are as follows:

\begin{figure}[h!]
\centering 
\includegraphics[width=0.45\textwidth]{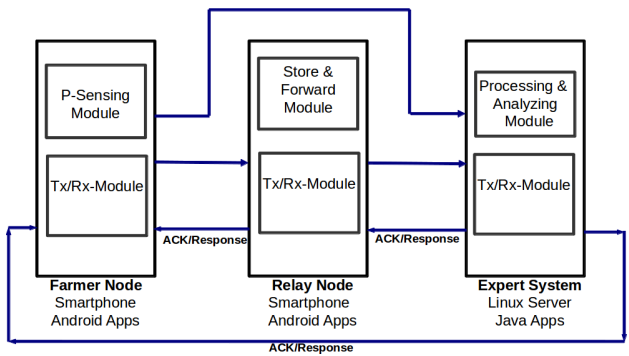}
\caption{Building Blocks - DTN-based Architecture}  
\label{soln_BB}\vspace{-2mm}
\end{figure}

{\bf{Farmer Node}}: In the present setup of the mKRISHI{\textsuperscript{\textregistered} framework, farmer node is an Android Application (App) targeted for users with Android devices having OS version not less than 3.0 and basic configurations like camera, Global Positioning System (GPS) and WiFi. The application empowers the users, farmers, situated in remote locations to send ground truth in the form of 'events' that fundamentally consists of an image of the situation or surrounding, predefined textual labels and/or a voice clip. In such cases where a particular farmer has a basic mobile phone instead of smartphone, he can still any other farmer's smartphone with his own authentication to upload his queries. In this case, there will be two different identities for the farmer - one physical identity, i.e., his own phone number, and one logical identity, which is the phone number of the smartphone he uses for log-in. These events are required to be uploaded to the server over the Internet so that users can 
receive advices or suggestions. The application allows users to store these events, for later submission, in case of poor or no network connectivity. While uploading, the system follows DTN approach to relay the data to an agent or relay node who can reach to a good network zone. 

{\bf{Relay Node}}: It employs the aforementioned farmer node Android App with an additional privilege of aggregating queries from farmer nodes which are incapable of uploading events. The relay node is required to press a button in the application to create a  WiFi Protected Access 2 (WPA2) secured WiFi Hotspot in the region. A limited set of farmer nodes (based on device hardware), simultaneously, could connect to the relay node and send their events to it. The relay node would then upload all the events to the server and the acknowledgement is sent to the respective farmer nodes.

{\bf{Server or Expert System}}: It is is a Linux powered, 16 core Xeon processor rack with 32GB RAM, hosted in a virtualized environment and capable of handling huge loads by applying standard techniques of distributed computing. It hosts the mKRISHI{\textsuperscript{\textregistered} server application which handles the 'events' coming from remote locations. The application is designed to maintain data integrity and has the capacity to send instant acknowledgements to the farmer nodes. 

Communication between the Farmer node and the Relay node is realized through low powered ad-hoc D2D communication, such as WiFi Direct, WiFi Hotspot and Bluetooth, etc. It uses the licence free spectrum (ISM band) and self-initiated discovery and signalling techniques for the D2D communication initialization. At present, it uses a very low power (fixed) for the D2D communication; transmission power can be further reduced using sophisticated signalling techniques.

\section{Deployment and Evaluation}\label{s:deployment}
To evaluate our proposed architecture, we have extended the mKRISHI{\textsuperscript{\textregistered} framework and included the DTN functionalities. The system model is implemented to work seamlessly under all communication scenarios. The details of our implementation framework is explained as follows:

\subsection{Scanning of Access Networks}
This module is implemented in the farmer node as well as in the relay node using Android App. Once the event is sensed, the farmer node creates a query message and scans for an access network to communicate the same to the central expert system. The communication can be direct through the cellular network or indirect through the mobile relay node.

\subsection{Communication Modes}
Based on the availability of access networks we have implemented two modes for access network selection and communication. Mode-1 belongs to direct communication whereas Mode-2 is Relay-based communication. In this paper we focus on Mode-2 communication only.

\begin{itemize}
\item \textit{Mode-1(a) - Direct:} When the Cellular or WiFi connectivity is available at the farmers' end, it can directly transmit the query to the central expert system.

\item \textit{Mode-1(b) – Direct Periodic (P-Direct):} Under this mode, the application at the farmers' end periodically scans the signal strength of the cellular network. Upon getting good connectivity, it can transmit the query through the available cellular network. This can be applicable in scenarios where the connectivity varies between good and bad within time periods in minutes. 

\item \textit{Mode-2 – Periodic DTN:}  This can be applicable in scenarios where the connectivity varies within time periods of hours. In such situations, if the channel is bad then most likely it will remain same for a long time. Therefore, for such situations a relay agent with a DTN enabled relaying application module will visit the farmer place and collect farmers' queries to its own device and then relay the collected queries to the server. The relay node can also collect the ACK/responses for the queries from the server and deliver to the farmer nodes through this communication mode. In this mode, the query won't be deleted from the farmer node till an ACK/response is received from the server. In case of no ACK/response received from either the server (directly) or through the relay node within a pre-define time frame, the farmer node will restart the access scanning process for re-transmission. 

\end{itemize}

\subsection{Relay Node Communication}
% The Relay node employs the aforementioned mKRISHI{\textsuperscript{\textregistered} Android application with an additional privilege of aggregating queries from farmer nodes who are not able to upload the queries due to poor network connectivity. 

For query collection from the farmer nodes, the relay node creates a WPA2 secured WiFi Hotspot in the region. Farmer nodes can register with this WiFi Hotspot and send their queries to the relay node. Once registered, the farmer's authentication details are saved in the relay node and need not to register again with the same relay node. In this way, secure D2D communication is realized between the farmer and the relay node. The relay node would then transmit all the queries to the server using its own network or by physically moving to a available network zone. Multi-hop relaying is also possible but currently out of the scope of this paper. Similar to the forward direction flow of queries, the relay node can also collect the ACK/responses for the queries from the server and deliver to the farmer nodes within a permissible time bound.

\subsection{Reverse flow of ACK/Response}
The ACK/response can be sent directly to the farmers using Short Message Services (SMS) or Hyper Text Transfer Protocol (HTTP) based messaging. In case of non-availability of network ACK/response can also be sent through designated relay nodes. In case of the query came from a farmer who has two identifications as physical and logical, the ACK/Responses will be routed to the logical identity only and it is the job of the farmer to logging into the smartphone and retrieve its message. The forward and reverse flow of message between the farmer node and the server can also happen using different relay nodes. To enable this multiple relay node options, the server maintains a map of relay nodes and their registered farmer nodes. Note that, Responses are required to be received by the farmer within a specified time period ($T_r$) starting from the query generation time (24-hour in this paper). Else, the farmer node discards the query; new queries can be generated again. Similar to the Responses, the ACKs are also 
to be received by the farmer node within a smaller time period $T_d$ ($T_d << T_r$). It is generally assumed that relay node can go online fairly quickly ($< T_d$) since it is mobile. To take care of the situation that the relay node due to some exigency is unable to connect to the server within this time period, we are having a timeout and a re-transmission to a possibly different relay node to ensure reliable message delivery to the server.

We now explain the control flow of all the events required for the end-to-end communication between the farmer node and the server. Fig. \ref{soln_timing} illustrates the different phases and modes of the communications. Under Mode-1 communication queries can be sent directly to the server node and ACKs/Responses can be delivered to the farmer node directly. Upon receiving the ACK the query associated with the event is deleted from the farmer node. In Mode-2, two cases are possible. In the first case, i.e., when the relay node is on-line, both the query as well as the ACK/Response can be received before the expiry of the desired time; event can be deleted after the ACK is received at the Farmer node. Note that, the ACKs/responses can also be received directly using SMS. In the second possible case, i.e., when the relay node is off-line, even after collecting the queries, the relay node may not be able to transmit the queries to the server immediately due to network issues. In this case, the relay node needs 
to physically move to a network zone to transmit the queries. This can lead to delay in receiving the ACK/response. Hence timeout can occur at the farmer node resulting in retransmission of the queries (and possible deletion of the query). Note that, retransmission of the farmer queries can be made through another relay node; multiple retransmission attempts can be made till the Response timeout.

\begin{figure}[h!]
\centering 
\includegraphics[width=0.48\textwidth]{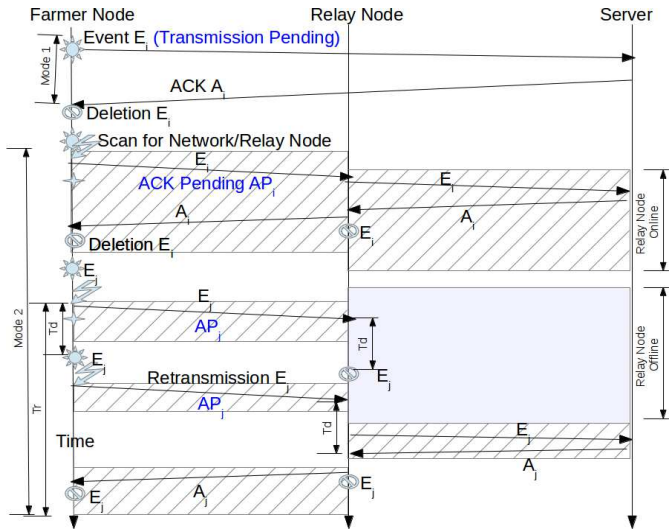}
\caption{Timing Diagram - DTN-based Solution}  
\label{soln_timing}\vspace{-2mm}
\end{figure}

%\subsection{mKRISHI\textsuperscript\textregistered Console Application}
\subsection{Proposed DTN-based Server Application} 
As discussed before, Android Apps are being designed for the farmer and relay nodes. Appropriate Java-based applications are also designed at the server end. These applications provide necessary platforms to visualize the queries, analyze them and communicate back responses or general information (advices/alerts) to the selected farmers (unicast or multicast). An SMS Gateway has been interfaced with the server in order facilitate the communication of ACKs, alerts, or responses in the form of the SMSs. 

Fig. \ref{soln_apps} shows the screen-shots of the Android Apps designed and implemented on the Farmer and the Relay nodes. These are simple and easy to use applications, which can be used by the farmers without much difficulty. Local language support is also being provided in the Apps, such that they can be adapted easily. Appropriate care has been taken by keeping security and privacy settings of smartphones intact while installing these Apps. 

\begin{figure}[h!]
\centering 
\includegraphics[width=0.48\textwidth]{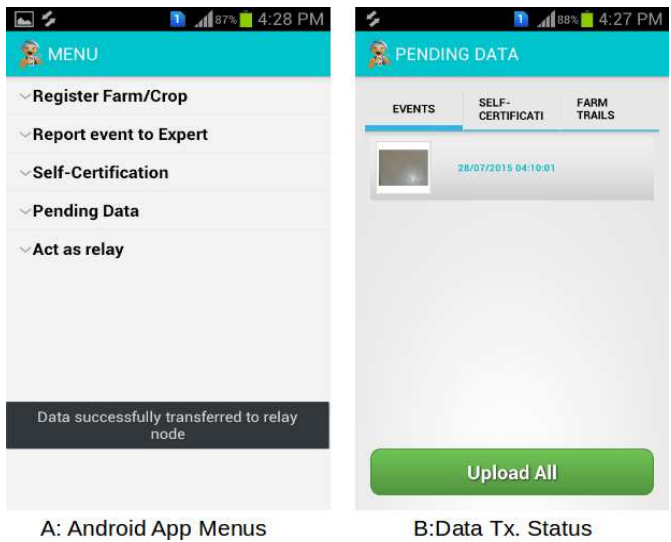}
\caption{Android-based Deployment}  
\label{soln_apps}\vspace{-2mm}
\end{figure}

\subsection{Practical Deployment}
At present laboratory prototypes are tested for (i) end-to-end data communication, (ii) ACK delivery through SMSs and through data communications, (iii) event creation, deletion, transmission and re-transmission, (iv) effect of D2D communication on interference and it's impact on throughput, (v) scalability test for the maximum number of Farmer nodes one Relay node can support, (vi) privacy and security threats which can result out of the relay-based communication, etc. Post laboratory tests, practical deployment will be considered in the Tea gardens of Assam followed by other areas in the North-East, Andhra Pradesh and Maharashtra.

\section{Conclusions}\label{s:conclusion}
In this paper, we have observed the need of a delay tolerant framework for the remote farmers with poor network connectivity and have proposed a novel DTN-based architecture for end-to-end communication. In our model, mobile relay node visits the poor channel farmer nodes, collects their queries and transmits to the server. The implementation of the DTN-based framework for first time in any Rural Participatory Sensing application has been presented in this paper. Simultaneous transmission of data through WiFi Direct and Cellular is also possible and can be tried out in future. Further, the estimation of the relay users availability at a certain location can help us to trigger other participatory sensing applications. This will enable the sensing operation just before the relay nodes are expected to arrive in communication range of the farmer nodes. We believe that this model can benefit the farming community to a large extent and can provide Digitization of the Agriculture domain in India.

\section*{Acknowledgement}
We thank Arun Kumar A. V. and Narendra N. of TCS Research, India for their support in the development work.

\bibliographystyle{IEEEtran}
\bibliography{allRefDTN}
\end{document}